\documentclass[12pt,preprint]{aastex}
\usepackage{txfonts}

\shorttitle{The CSTAR Point Source Catalog}

\shortauthors{Zhou et al.}

\def\DA{Dome\,A}
\def\DC{Dome\,C}

\begin{document}
%\slugcomment{AJ, in press}
\title{The First Release of the CSTAR Point Source Catalog from Dome A, Antarctica}

\author{Xu Zhou\altaffilmark{1,5}, Zhou Fan\altaffilmark{1}, 
  Zhaoji Jiang\altaffilmark{1,5},   M.C.B. Ashley\altaffilmark{4},  
  Xiangqun Cui\altaffilmark{2,5},  Longlong Feng\altaffilmark{3,5}, 
  Xuefei Gong\altaffilmark{2,5}, Jingyao Hu\altaffilmark{1,5},
  C. A. Kulesa\altaffilmark{8},  J.S. Lawrence\altaffilmark{4,6,12}, 
  Genrong Liu\altaffilmark{2}, D.M. Luong-Van\altaffilmark{4}, 
  Jun Ma\altaffilmark{1}, A. M. Moore\altaffilmark{9}, 
  Weijia Qin\altaffilmark{7}, Zhaohui Shang\altaffilmark{10},
  J.W.V. Storey\altaffilmark{4}, Bo Sun,\altaffilmark{7},
  T. Travouillon\altaffilmark{9}, C. K. Walker\altaffilmark{8},
  Jiali Wang\altaffilmark{1,5}, Lifan Wang\altaffilmark{3,5},
  Jianghua Wu\altaffilmark{1}, Zhenyu Wu\altaffilmark{1}, 
  Lirong Xia\altaffilmark{2}, Jun Yan\altaffilmark{1,5},  
  Ji Yang\altaffilmark{3}, Huigen Yang\altaffilmark{7}, 
  Xiangyan Yuan\altaffilmark{2,5}, D. York\altaffilmark{11},
  Zhanhai Zhang\altaffilmark{7},  Zhenxi Zhu\altaffilmark{3,5}}

\altaffiltext{1}{National Astronomical Observatories, Chinese Academy of
  Sciences, Beijing, 100012, P. R. China;\\ zhouxu@bao.ac.cn}
\altaffiltext{2}{Nanjing Institute of Astronomical Optics and Technology, Nanjing 210042, China}
\altaffiltext{3}{Purple Mountain Observatory, Nanjing 210008, China}
\altaffiltext{4}{School of Physics, University of New South Wales, NSW 2052, Australia}
\altaffiltext{5}{Chinese Center for Antarctic Astronomy}
\altaffiltext{6}{Current address: Macquarie University \& Anglo-Australian Observatory}
\altaffiltext{7}{Polar Research Institute of China, Pudong, Shanghai 200136, China}
\altaffiltext{8}{Steward Observatory, University of Arizona, Tucson, AZ 85721}
\altaffiltext{9}{Department of Astronomy, California Institute of Technology, Pasadena, CA 91125}
\altaffiltext{10}{Tianjin Normal University, Tianjin 300074, China}
\altaffiltext{11}{Department of Astronomy and Astrophysics and Enrico Fermi Institute, University of Chicago, Chicago, IL 60637}
\altaffiltext{12}{Anglo-Australian Observatory, Epping, NSW 2121, Australia}

\begin{abstract}
  In 2008 January the 24$^{\rm th}$ Chinese expedition team successfully deployed 
  the Chinese Small Telescope ARray (CSTAR) to \DA, the highest point on the 
  Antarctic plateau. CSTAR consists of four 14.5cm optical telescopes, each 
  with a different filter (g, r, i and open) and has a 
  $4.5^{\circ}\times 4.5^{\circ}$ field of view (FOV). It operates robotically 
  as part of the Plateau Observatory, PLATO, with each telescope taking an
  image every $\sim$30 seconds throughout the year whenever it is dark.  During
  2008, CSTAR \#1 performed almost flawlessly, acquiring more than 0.3 million 
  i-band images for a total integration time of 1728 hours during 158 days of observations. For each image taken under 
  good sky conditions, more than 10,000 sources down to $\sim16^{\rm th}$ magnitude
 could be detected. We performed aperture photometry on all the sources in 
 the field to create the catalog described herein. Since CSTAR has a fixed pointing 
 centered on the South Celestial Pole (Dec $=-90^{\circ}$), all the sources 
 within the FOV of CSTAR were monitored continuously  for several months. 
 The photometric catalog can be used for studying any variability in these
 sources, and for the discovery of transient sources such as supernovae, gamma-ray bursts and 
 minor planets.
\end{abstract}


\section{Introduction}
\label{Introduction.sec}

The Antarctic Plateau has great potential for astronomical
observations as it is extremely cold and dry and has a calm  and
tenuous atmosphere---attributes that are particularly favourable for
optical, infrared and {\bf submillimeter} observations.  For
photometry, there is the  advantage of reduced scintillation as a
result of the decreased high-altitude turbulence above the Antarctica
plateau. \citep{ken06}. In addition, there  is the {\bf possibility} of
long, continuous observations uninterrupted by the usual  diurnal
cycle, giving access to a time-series regime otherwise only available
from space \citep{ken06a, mos07, rau08}. 

Astronomers have been interested in the Antarctic plateau for over
thirty years (see \citet{ind05} for a historical account). The first
stellar photometry from Antarctica was conducted at the South Pole by
\citet{tay88} in the late 1980s.  More recently, Strassmeier and
colleagues \citep{str08} have conducted continuous time-series
observations from \DC\ with sIRAIT, a small f/12 optical
telescope. The ASTEP South project has reported 1592 hours of
observations of the South Celestial Pole from \DC\ during 2008 using a
10 cm refractor and $4096\times4096$ pixel CCD camera \citep{cro09}. 

At optical wavelengths most of the attention is now focussed on the Concordia 
station at Dome C, where excellent cloud-cover statistics, low free-atmosphere 
seeing and a relatively thin turbulent boundary layer have been measured, as 
summarised in e.g., \citet{sto05}. For example, \citet{law04} investigated the 
seeing at \DC\ and reported that the median seeing above the boundary layer 
is $0.27''$, and for 25\% of the time is as low to 0.15$''$. Thus, for some observations even a small
telescope at \DC\ can be as effective as a much larger one at the best 
temperate observatories.

\DA\ is located at longitude $77^{\circ}06'57''$E, latitude
 $80^{\circ}25'08''$S, and is 1271 km directly inland from the Chinese Zhongshan
Station. The elevation of \DA\ is 4093 meters above sea level, the temperature
is quite low (sometimes below $-80 ^{\circ}$C), the surface wind speeds are
even lower than those at \DC, and the precipitable water vapor is extraordinarily low \citep{kul09}. 
The seeing at \DA\ from above the boundary layer may be better than that 
at \DC, and the boundary layer itself should be thinner---perhaps as low as
15 meters for much of the time.

Recently, \citet{sau09} compared Domes A, B, C and F, and Ridges A and
B from the point of view of cloud cover, boundary layer thickness,
auroral emission, free-atmosphere seeing, precipitable water vapor and
surface temperature---and concluded that \DA\ is possibly the best
site on earth currently being used for astronomical observations.

Because of the potential opportunities offered by \DA\ for astronomy and 
other scientific studies, an permanent station, {\it Kunlun station}, was 
established at \DA\ by the 25$^{\rm th}$ Chinese expedition team in 2009 January.

In early 2007, Chinese astronomers began the development of CSTAR, with 
an extensive testing program (Zhou et al. 2009). In November of that year, the 
CSTAR system was shipped to Antarctica as part of the Plateau Observatory 
(PLATO) \citep{law08, yang09}, and commissioned at \DA\ in 
2008 January. CSTAR began observations from \DA\ in 2008 March, as soon as the 
sky became sufficiently dark. During the year, we were typically able to return 
one image per day and roughly one third of the photometric catalog (less 
than 3000 bright stars) via Iridium satellite. CSTAR worked well until early 
August, when the PLATO power system shut down. The 25$^{\rm th}$ Chinese expedition
team returned to \DA\ in 2009 January, retrieving all the data including 
images and catalogs that were stored on a large hard disk. The \DA\ data are
valuable as they have many potential uses, from site characterisation to the 
study of astronomical sources such as variable stars, supernovae, gamma-ray
bursts and extra-solar planets. A separate paper analysing the weather 
statistics of the \DA\ site from CSTAR data is in preparation \citep{zou10}.

This paper is organized as follows. A description of CSTAR
and the data reduction method are presented in \S 2. The
observations from \DA\ are described in \S 3. In \S 4, we describe the
data processing of these observations to produce
the catalog. In \S 5, we discuss the photometric accuracy of the
catalog. A final summary is presented in \S 6.

\section{CSTAR}

\subsection{The instrument}
The CSTAR program is conducted under the auspices of the Chinese Center for
Antarctic Astronomy. The telescopes were built by the Nanjing Institute of 
Astronomical Optics \& Technology (NIAOT) \citep{yuan08}, while the hardware 
and software of the data acquisition system were developed by the National 
Astronomical Observatories, Chinese Academy of Sciences (NAOC) \citep{zhou09}.
CSTAR consists of four 14.5cm Schmidt telescopes, each with a different 
filter: g, r, i and open. The FOV is $4.5^{\circ}\times 4.5^{\circ} ($20 deg$^2$).
Each telescope is equipped with an 
Andor DV435 1k$\times$1k CCD, giving a pixel size of about 15$''$ in the sky. 
A 750 GB hard disk is used as the main storage, with all the software including
the Windows XP operating system installed in a 4 GB Compact Flash memory. 
An 8 GB solid state disk is used in each computer to back up the most important
data. The computer system can work at temperatures down to $-30^{\circ}$C, and 
is installed inside the PLATO instrument module where it is kept well above 
this temperature by the PLATO thermal management system \citep{luo08}. The 
CCD cameras and electronics are installed outside on the CSTAR telescopes,
where the ambient temperature can be lower than $-80^{\circ}$C. 

Before its deployment to \DA, the CSTAR data acquisition system was tested 
at Kalasu on the high plateau of Pamir in China, close to Tajikistan. 
At an elevation of 4450 meters and with 
temperatures that plunged to $-18 ^{\circ}$C during the tests, Kalasu provides
an excellent environment for pre-deployment testing of astronomical
instruments for \DA.  The complete CSTAR system was also tested at the NAOC's
Xinglong observatory at the beginning of September of 2007. Details of these 
tests are presented in \citet{zhou09}.

In \citet{zhou09} we also describe the data acquisition procedure and
automatic photometric pipeline used to perform aperture photometry on
all the point sources in our images. The apertures for the photometry
are 3, 4 and 5 pixels in radius and the inner and outer radius of the
sky annulus is  10 and 20 pixels, respectively. 

\begin{figure*}
\center
\includegraphics[scale=0.4,angle=0]{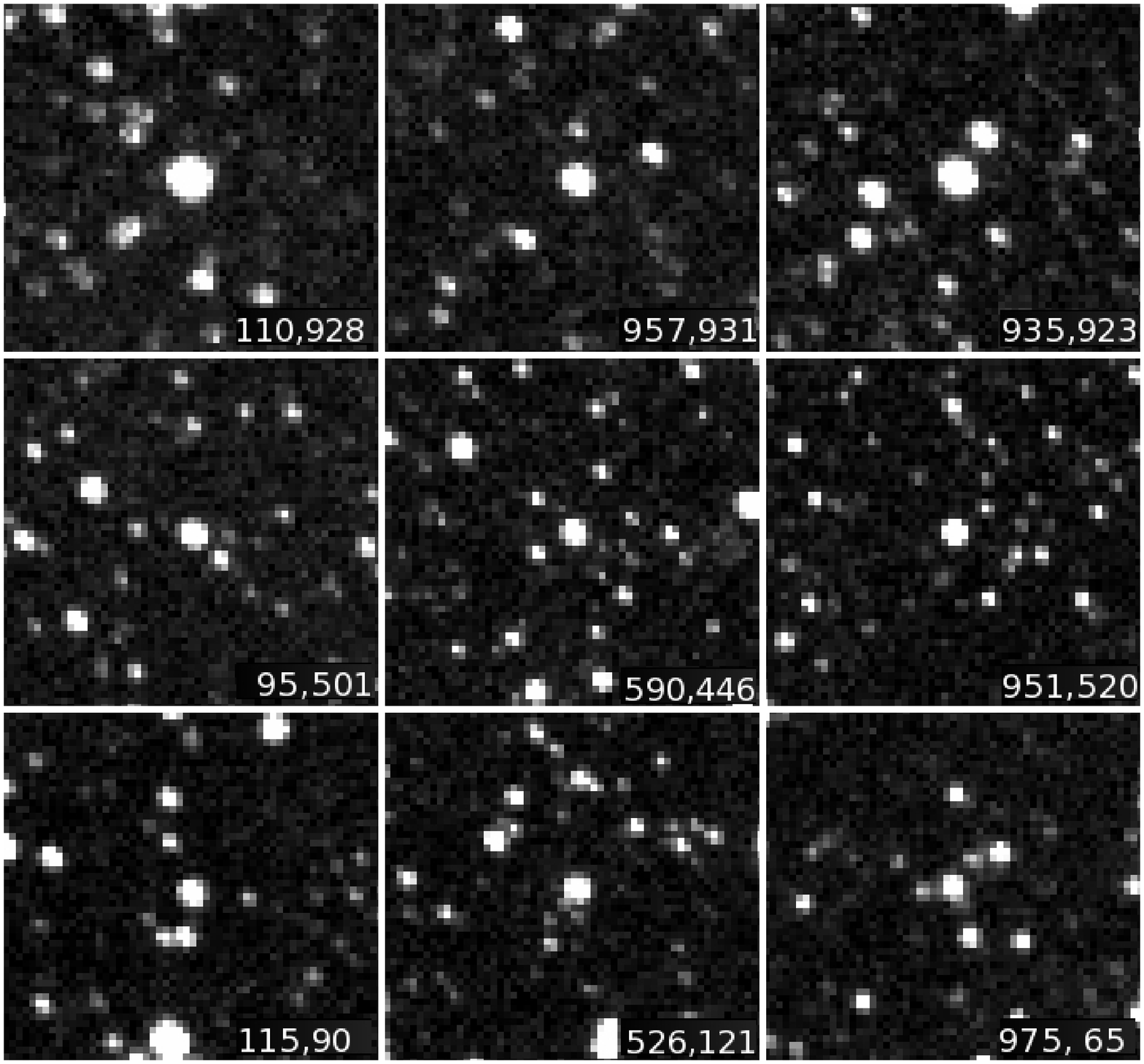}
\caption[]{$50\time50$ pixel sub-images of the image A5CH5029.fit are 
           used to show the PSF located in the center, four corners and four sides of the 
           field of view}
\label{fig:9}
\end{figure*}

The typical FWHM of
the point spread  function (PSF) of a stellar image taken by CSTAR\#1
is between 1.5 and 2.5  pixels.  {\bf The spot diagram under laboratory
conditions, that is $20^{\circ}$C  and 1 atm pressure, is shown in
\citet{yuan08}.  The optical system may have suffered some misalignment from vibration
during the almost 1300km sled traverse from Zhongshan Station to Dome A.
It is also possible that there were some optical alignment changes due
to the low ambient temperature at Dome A, although this was allowed for in
the design. An image from CSTAR\#1, A5CH5029.fit, is shown in Figure~\ref{fig:9}
to illustrate the realized optical quality at Dome A, where $\sim90\%$ of the light energy is encircled in
2 pixels.}

\section{Observations at Dome A}
\begin{figure*}
\center
\includegraphics[scale=0.5,angle=0]{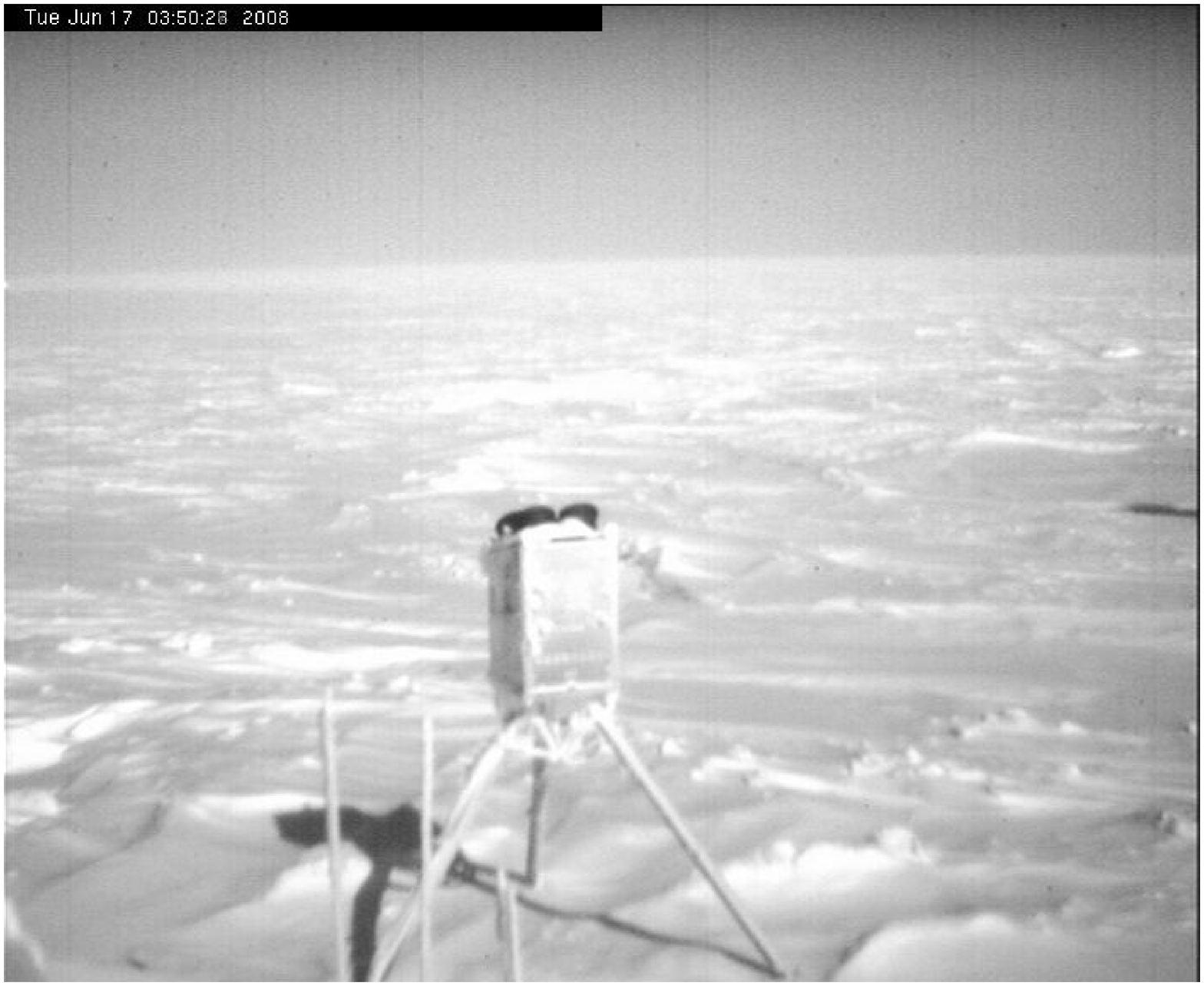}
\caption[]{CSTAR during a mid-winter full moon at Dome A, Antarctica,
  on 2008 June 17.  This image was taken by a camera on the roof of
  PLATO's Instrument Module \citep{law08}. The four CSTAR telescopes are
  in the box on the tripod, pointing away from the camera. For scale, the
  top of the CSTAR box is about 1.8m above the snow.}
 \label{fig:10}
\end{figure*}

\begin{figure*}
\center
\includegraphics[scale=0.7,angle=0]{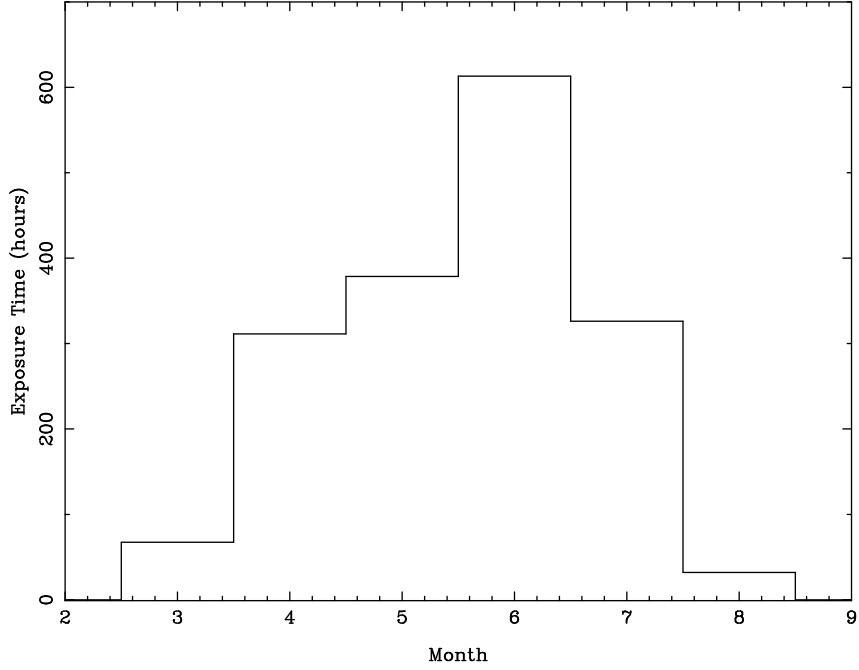}
\caption[]{The integrated CSTAR exposure time for each month during 2008.}
 \label{fig:1}
\end{figure*}

CSTAR observed robotically at \DA\ from 2008 March 4 until 2008 August
8 (see Figure \ref{fig:10}),  taking an image with each telescope whenever it was
dark  enough to do so. The exposure time was changed from 20 sec to
30 sec on April 4.  The CCD is a frame transfer CCD, which allows us to operate
without a mechanical shutter. We readout the image after
each exposure  and perform a real-time data reduction for producing
an initial catalog.  A deadtime of $\sim$2.61 sec
results in times between exposures of 22.61 sec, or 32.61 sec 
after April 4. Images were not taken if the sky was too bright.
For a variety of technical
reasons only CSTAR\#1  produced images of good optical quality during 2008.
There are 16918, 50575, 68157, 110357, 58711 and 5749 of these
images from March, April, May, June, July and August,
respectively. The total exposure times were 67.4, 311.4, 378.7,
613.1, 326.2 and 31.9 hours, respectively, in these months, giving a
total exposure time of 1728 hours. We plot the distribution of
integrated exposure time for each month in Figure~\ref{fig:1}. The
integrated exposure time in June is the longest and is over one third
of the total exposure time, while the exposure time in August is the
shortest due to the shutdown of the PLATO power system.

\begin{figure*}
\center
\includegraphics[scale=0.7,angle=-90]{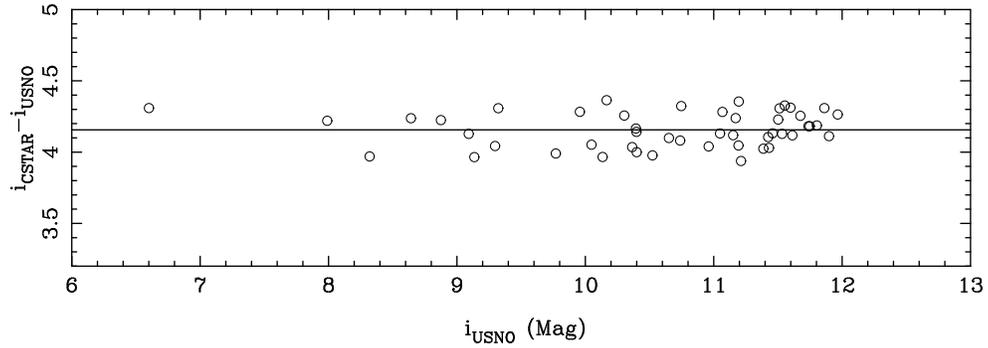}
\caption[]{Comparison of the USNO-B1 i-band magnitudes and CSTAR instrumental 
  magnitudes for 48 stars. The average offset is 
  $\Delta i=4.16\pm0.12$\ mag.}
 \label{fig:2}
\end{figure*}

\section{Data Reduction}

We have developed a pipeline for image processing and photometry
that includes bias and flat-field correction, and then performs aperture
photometry \citep{zhou09}. This work was undertaken during testing CSTAR at
NAOC's Xinglong station. The quality of the images taken at \DA\ was 
significantly better than obtained at Xinglong due to the lower
CCD temperature and improved sky conditions.

\subsection{Flat fielding}
\label{sec:flat}

\begin{figure*}
\center
\includegraphics[width=85mm]{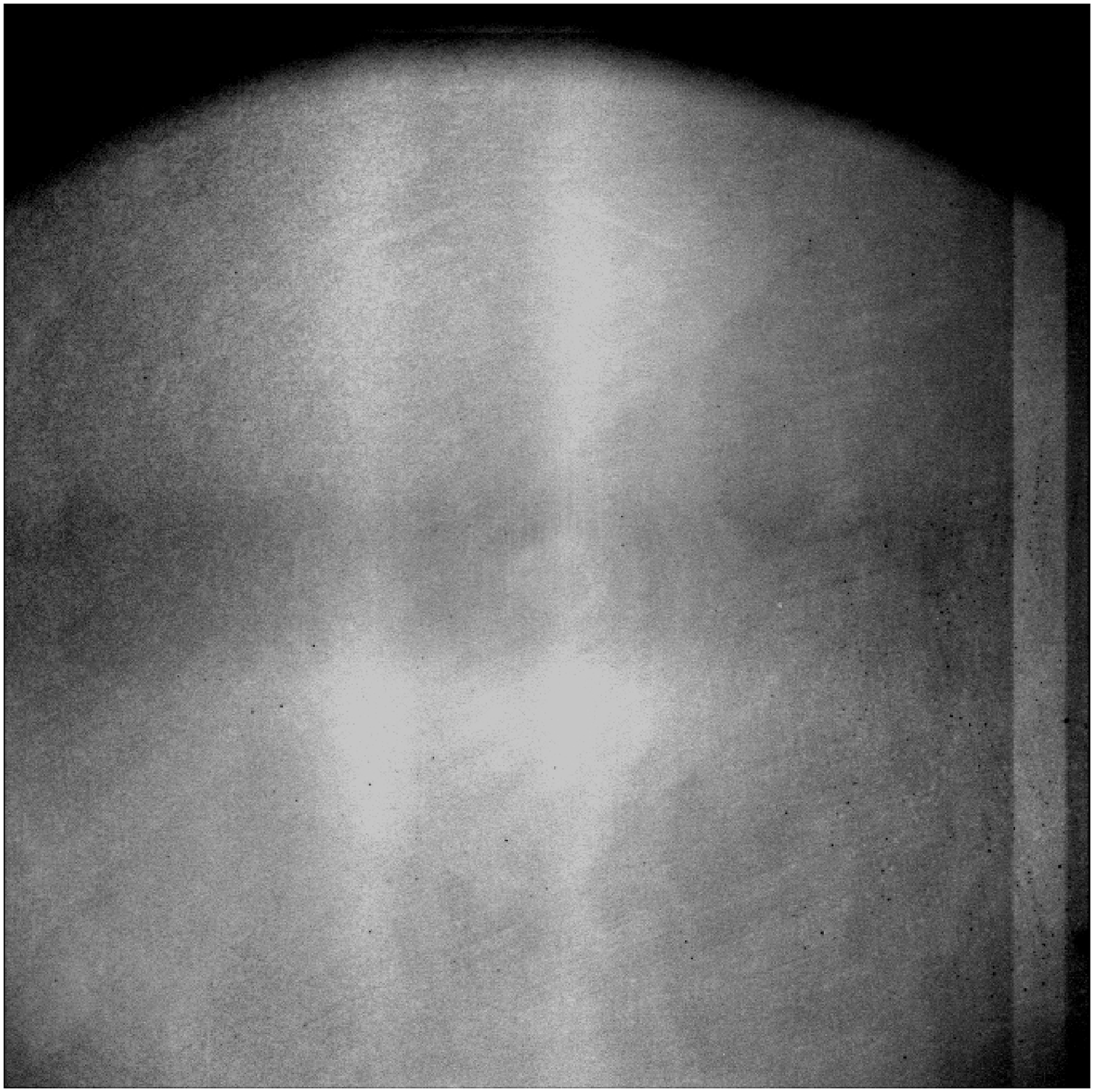}
\caption{The flat-field image used for CSTAR\#1.}
\label{fig:3}
\end{figure*}

{\bf We constructed a flat-field image, shown in Figure~\ref{fig:3},
  from a median of images taken with a high sky background, after
  removal of stars using sigma-clipping. To correct for variations in the flat-field
  over large spatial scales, we selected more than 50,000 images taken under
  conditions of good transparency and looked for systematic changes in the
  brightnesses of stars during their daily circular motion around the South Pole.
  This allowed us to create an additional ``residual flat-field'' correction that
  improved our photometric accuracy.}

%\begin{figure*}
%\center
%\includegraphics[width=85mm]{fig4.ps}
%\caption{The residual of flat-field correction of CSTAR\#1.}
%\label{fig:4}
%\end{figure*}

 {\bf For this first release of the catalog we have not checked the
   variation of the flat-field during the year, nor with CCD temperature. 
   The next release will
   include improved flat-fielding using images taken under photometric
   conditions with high sky background (i.e., during twilight or bright
   moon).}

\subsection{Absolute flux calibration}
\label{sec:cali}

{\bf An image, A5CH5029, was taken under relatively good photometric
  conditions (17:50:29 UT, 2008 May 5) and was used as a standard
  image for calibrating magnitude offsets. For each other image, we
  derived a single magnitude offset, to be applied to all the stars on
  the image, from the mean of the magnitude offsets of a selection of
  bright stars.}
 
The USNO-B catalog \citep{mon03} contains stellar magnitudes in the
optical passbands B1, B2, R1, R2 and I2 for over one billion objects
over  the whole sky. The catalog is complete down to $V = 21$~mag, with
a positional accuracy of $0.2''$ at J2000 and a photometric accuracy
of better than 0.3 magnitudes in the five colors. Since the USNO-B
catalog contains such well calibrated magnitudes of the point sources
in our observed field, we elected to use these objects for our flux
calibration.  Fortunately, \citet{mon03} have derived a formula for
transforming the USNO-B1.0 magnitudes to those appropriate to the
Sloan Digital Sky Survey (SDSS) filters, based on $\sim 450$ deg$^2$ of
SDSS Early Data Release. Since the CSTAR filters were chosen to be
very similar to the SDSS  filters, we can use this formula directly.
We determine
$\overline{i {\rm CSTAR}-i {\rm USNO}}=4.16\pm0.12$ with 48 field
stars.  The differences between CSTAR instrumental magnitude and SDSS
magnitudes  of these stars are shown in Figure~\ref{fig:2}. The
offset was used to transform the CSTAR instrumental magnitudes 
to SDSS i-band magnitudes. {\bf To reduce statistical measurement 
errors,  we only selected the brightest stars for the offset calibration. 
We plan to do further work on choosing well-calibrated standard stars
for our next run of the data reduction pipeline.}

\begin{figure*}
\center
\includegraphics[scale=0.7,angle=0]{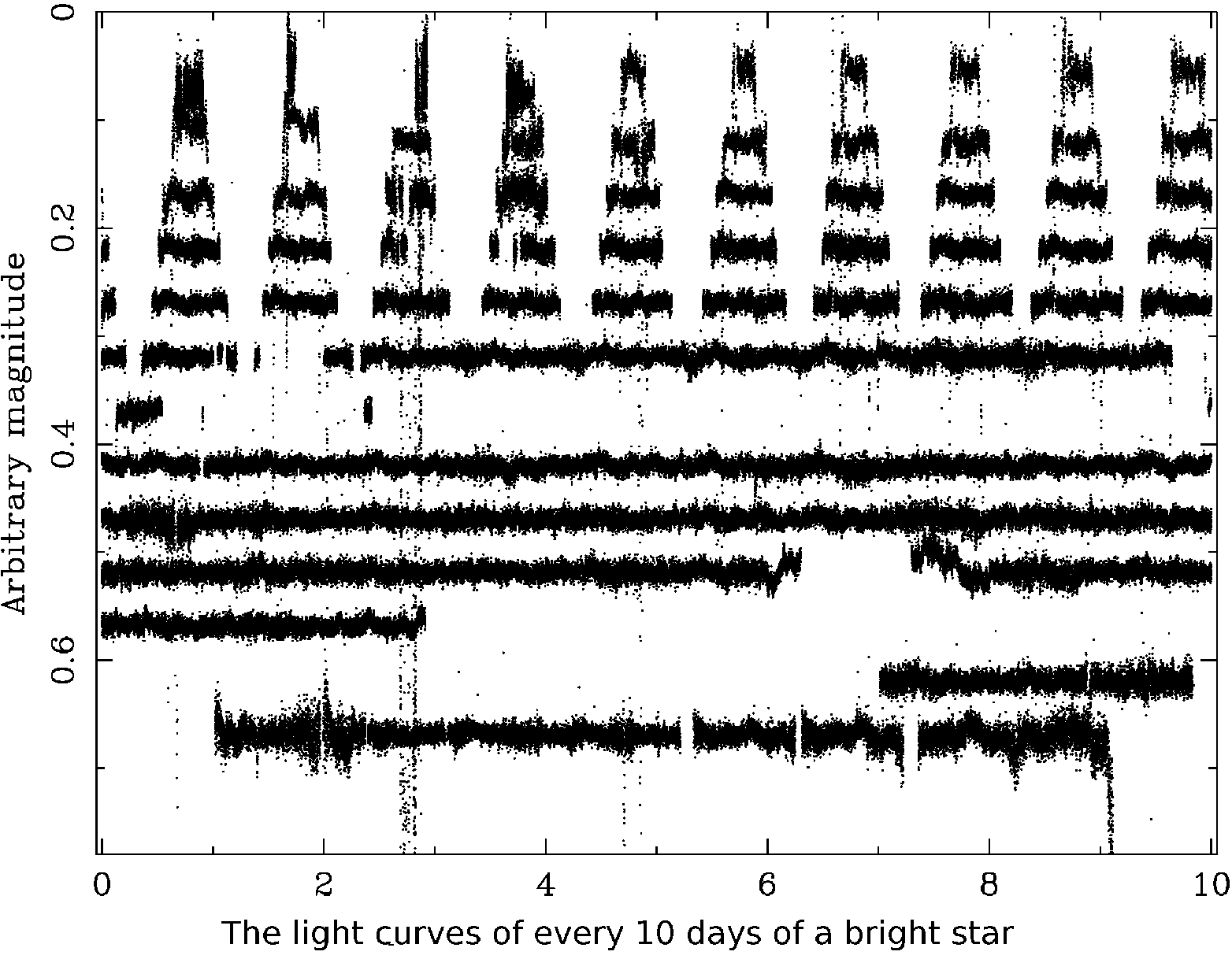}
\caption[]{The light curve of HIP 48752 = 
GSC 9518:379 at $09^{h}57^{m}43.3^{s}, -89^{\circ}47^{'}02.2^{''}$, a 8.2 mag
star. The gaps
at the top of the Figure are due to twilight.}
 \label{fig:5}
\end{figure*}

\subsection{Time calibration}

\begin{figure*}
\center
\includegraphics[scale=0.5,angle=0]{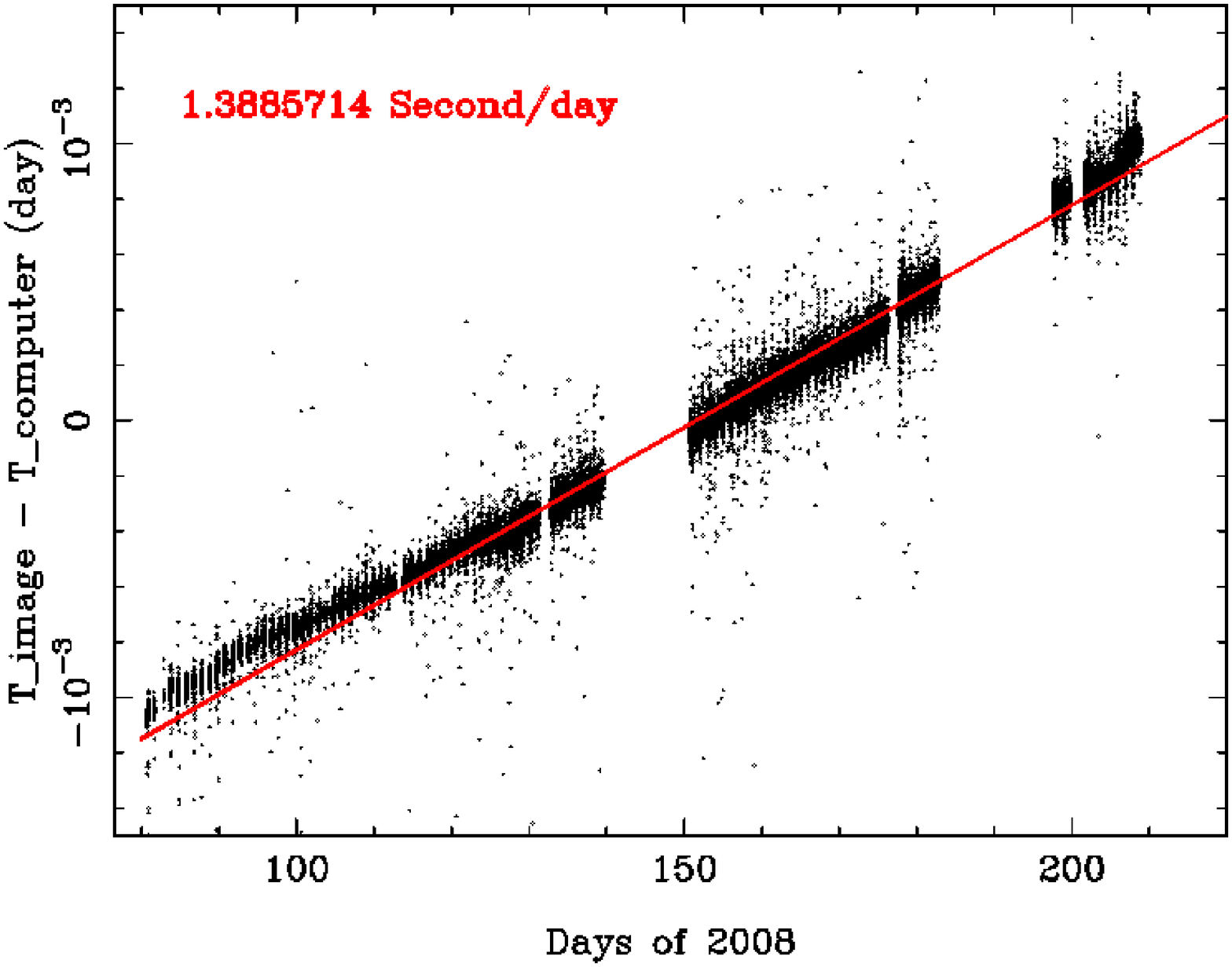}
\caption[]{The difference between computer time and the true local time 
  throughout the observational period. The true local time was obtained from 
  the star positions on the images.}
 \label{fig:6}
\end{figure*}

The CSTAR\#3 computer included a GPS receiver to maintain time
synchronisation  and this time was intended to be distributed to the
other CSTAR computers.  However, there was a communication problem
between CSTAR\#3 and the other  computers throughout the year.  As a
result, the timing of CSTAR\#1 ran  independently and at its own rate
for the entire observation period. While  this would normally create
an intractable data reduction problem---particularly when determining
the epochs of transient events and eclipses---in this case it is
easily corrected. CSTAR is fixed in position and points at the  South
Celestial Pole. Every star traces out a circle on the CCD, and the
position  of each star can be used as a clock. To identify the stars,
we calculate the  rotation and shift of every image relative to a
standard image using the positions of all the bright stars.  The mean
rotation angle relative to the standard image is then used to derive
the rate of drift of computer time. In this way,  precise timing can
be determined. We find that the computer clock typically runs
1.3885714 seconds/day slower than the real local
time. Figure~\ref{fig:6} shows the time difference between the
computer time and the real time obtained from the star positions
during a period of 140 days of observation. The computer time on 2008
January 25 should be correct, because the computer time was manually
set to GPS time (UT) on that date. In the  catalog we present the
corrected Julian time at the mid-exposure point of  every image. This
time should be accurate to {\bf a few seconds}. 

\subsection{Photometric accuracy}

Besides the statistical errors in the star and background brightness 
measurements, there are several systematic errors that may affect the derived 
stellar fluxes. The main systematic errors are:

1) Residual flat-field and bias correction errors. {\bf Light curves
  of all of the stars that had no significant variation were used to
  improve the flat-field correction map. However, a single flat-field
  and bias image were used for all observations, with no attempt to
  allow for temperature variations or instrumental drifts. This may
  have introduced systematic errors with each star tracing out a circle
  on the CCD.}

2) Point Spread Function (PSF) variation. The CSTAR telescopes have a large 
field-of-view, $> 20$ square degrees, making it impossible for the optical design to keep the PSF 
uniform over all parts of the image. Since we used a fixed aperture to measure 
the magnitudes of the stars, there will be some measurement error from the PSF 
variation depending on the star location within the image. 

3) Under-sampling. Each pixel of the CSTAR CCD is
15$''$, so the light from a star might fall on only one or two pixels, with
resultant problems from intra-pixel sensitivity variations.

4) Aurorae and thin cirrus clouds. These introduce an
inhomogeneity in the sky background, especially during full moon. 
In the case of cirrus, there will be a resultant variable 
extinction across the field-of-view. The photometric calibration can therefore 
differ from one star to another. The sky background for each star is also difficult
to estimate in the presence of cirrus.

The systematic photometric error exceeds other sources of error when observing bright 
stars. As an example, we plot the light curve of a bright star (HIP 48752 = 
GSC 9518:379) of i$\,=\,$8.2~mag at
($09^{h}57^{m}43.3^{s}, -89^{\circ}47^{'}02.2^{''}$) in Figure~\ref{fig:5}. 
In the Figure, the light curves for every 10 days, with an arbitrary magnitude
offset, are shown. From examining the best quality images, the RMS of the light 
curve variation is about 0.003 mag. We expect that the systematic error should 
be smaller than 0.003 mag.

To estimate the overall real measurement error for
each field  star in the image, we compared every image to a standard
image.  The standard image, A5CH5029, was taken under relatively good
photometric conditions (17:50:29 UT, 2008 May 5). By comparison of the
calibrated magnitudes of all the stars in the two images, we can
plot the RMS error as a function of the magnitude of
the stars. 

\begin{figure*}
\center
\includegraphics[scale=0.7,angle=0]{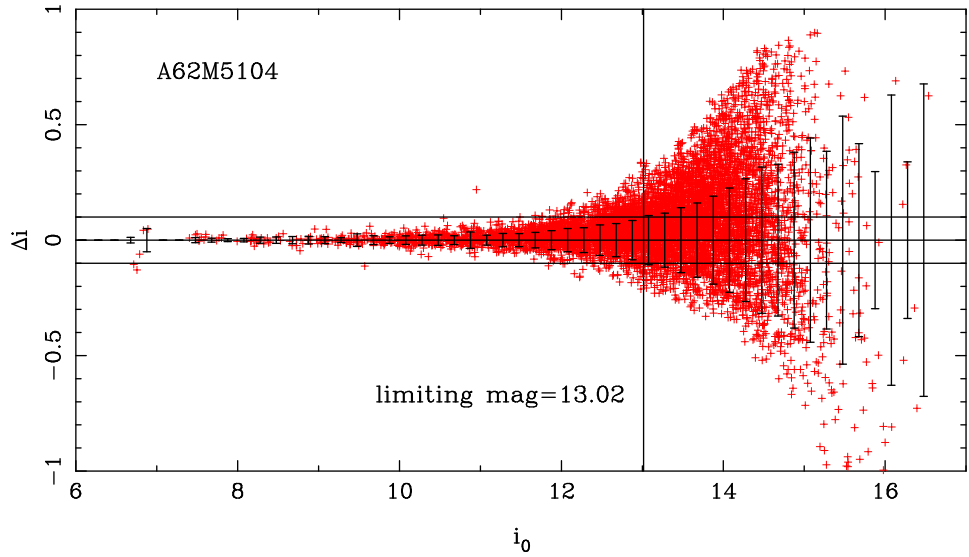}
\caption[]{Estimate of the photometric error for all the point sources
of A62M5104. The error bars represent the 1$\sigma$
errors. The two horizontal lines represent $\Delta$mag=0.1, while the vertical 
line shows the corresponding limiting magnitude of 13.02 at S/N=10.} 
\label{fig:7}
\end{figure*}

As an example we show in Figure~\ref{fig:7} the distribution of the
measurement difference in magnitude for every star obtained from
image A62M5104 and the standard image A5CH5029. We divided the
magnitude  range into small bins of 0.2 mag, then calculated the
$1\sigma$ dispersion  of the magnitude as the ``real measurement
error'' in each magnitude interval.  The error bars as a function of magnitude as shown in
Figure~\ref{fig:7} were then applied to all the  measurements in the catalog for that image.

We defined the magnitude limit of our images as the magnitude which has a 
$1\sigma$ RMS error of 0.1 mag, which corresponds to a S/N=10.

\begin{figure*}
\center
\includegraphics[scale=0.7,angle=0]{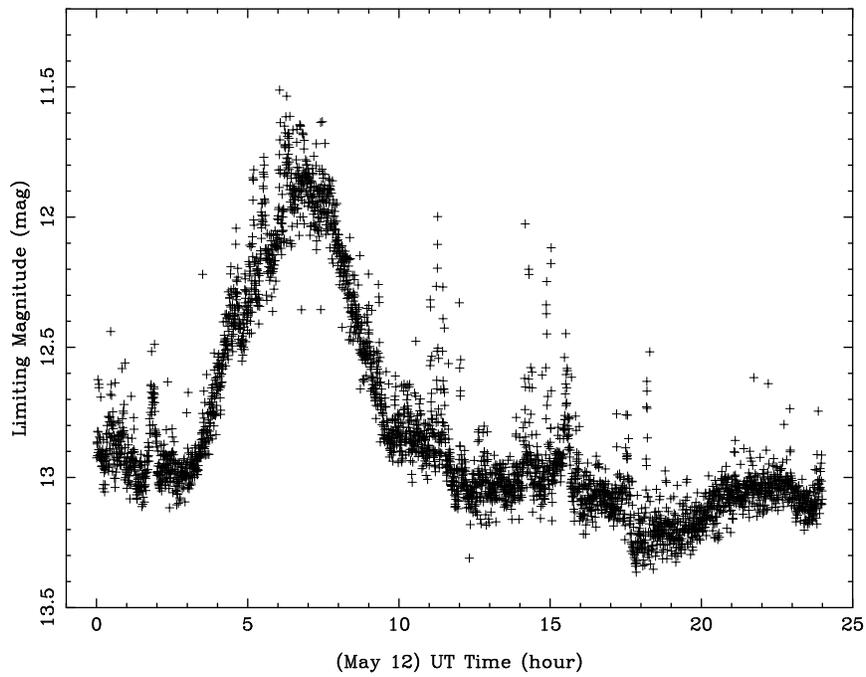}
\caption[]{The limiting magnitudes for one whole day (24 hours) of
images on May 12, 2008.}
\label{fig:8}
\end{figure*}

As a byproduct of estimating the measurement error, the magnitude 
limit is also obtained for each image. In Figure~\ref{fig:8} we plot 
the limiting magnitude distribution for all of
the images taken over the 24 hours during 2008 May 12. The limiting magnitude
changes with time because of variations in atmospheric extinction
and sky brightness.

The final output from our data reduction is a catalog of star magnitudes for each image.
The contents of the catalog are arranged as follow: 
The first two columns are RA and DEC, respectively; the following
columns are the magnitudes and errors in aperture photometry obtained with
radii of 3, 4, and 5 pixels respectively. In the header of the
catalog some additional information is provided:
the CCD temperature, the date and corrected time (UT) at the exposure midpoint, exposure time
(seconds), sky brightness (ADU), filter number and number of sources found in the image.

\begin{table*}
\begin{center}
\caption[]{Photometry of several sources in three different apertures. The catalog header is: ``$-$59 2008 Jun 02 22:50:42.20 20 i 10398 154.954086''
\label{tab:cata}}
\begin{tabular}{c|cc|cc|cc|cc}
  \hline\hline \noalign{\smallskip} \multicolumn{1}{c}{Number}&\multicolumn{1}{c}{RA}&\multicolumn{1}{c}{DEC} & \multicolumn{1}{c}{$M 1$} &\multicolumn{1}{c}{$\sigma 1$} &\multicolumn{1}{c}{$M 2$} &\multicolumn{1}{c}{$\sigma 2$}&\multicolumn{1}{c}{$M 3$} &\multicolumn{1}{c}{$\sigma 3$} \\
  \multicolumn{1}{c}{}&\multicolumn{1}{c}{(J2000)}&\multicolumn{1}{c}{(J2000)} &\multicolumn{1}{c}{(r=3~pixel)}&\multicolumn{1}{c}{} &\multicolumn{1}{c}{(r=4~pixels)}&\multicolumn{1}{c}{} &\multicolumn{1}{c}{(r=5~pixels)}&\multicolumn{1}{c}{} \\ \hline \noalign{\smallskip}
  277 & 23:23:46.274 & -89:25:17.81& 11.095 & 0.022& 11.011&  0.022& 10.838&  0.025\\
  278 & 10:43:24.023 & -88:42:00.78& 11.099 & 0.026& 11.048&  0.022& 11.013&  0.022\\
  279 & 16:13:39.187 & -87:44:30.11& 11.100 & 0.026& 11.030&  0.022& 11.014&  0.022\\
  280 & 14:09:08.706 & -89:07:12.63& 11.100 & 0.026& 11.035&  0.022& 10.987&  0.022\\
  281 & 13:46:15.127 & -88:26:01.94& 11.100 & 0.026& 11.013&  0.022& 10.885&  0.025\\
  282 & 17:54:27.175 & -89:42:21.70& 11.103 & 0.026& 11.065&  0.022& 11.032&  0.022\\
  \noalign{\smallskip}\hline
\end{tabular}
\end{center}
{NOTE. The header parameters are decoded as: (1) $-59$: CCD temperature in Celcius (2) 2008 Jun 02: date 
  (3) 20: exposure time in seconds (3) 10398: the number of sources detected 
  in the image (4) 154.954086 day of the year during 2008. The catalogs can be 
  downloaded from National Astronomical Observatories Science Data 
  Center, Chinese Academy of Science at http://archive.bao.ac.cn/en/cstar.
}
\end{table*}

\section{Summary and catalog availability}
The CSTAR Point Source Catalog first release contains 1728 hours of i-band photometry data 
taken by CSTAR\#1 at Dome A, Antarctica, between 2008 March 4 and 2008 August 
8. The data are from a fixed field-of-view of $4.5^{\circ}\times 4.5^{\circ}$ 
centered on the South Celestial Pole, with an image taken approximately every 30 s. Aperture photometry was used to derive the magnitudes of each 
of the 10,000 stars that were typically identified in each of the 300,000 images. The 
data have been flux calibrated using 48 standard stars to link to the USNO-B1.0 
photometric system. The CSTAR catalog is available at http://archive.bao.ac.cn/en/cstar.

\section*{Acknowledgments}

This study has been supported by the Chinese National Natural Science
Foundation through grants 10873016, 10803007, 10473012, 10573020,
10633020, 10673012, and 10603006, and by the National Basic Research
Program of China (973 Program), No.~2007CB815403.  This research is
also supported by the Chinese PANDA International Polar Year project,
the Polar Research Institute of China (PRIC), and the international
science and technology cooperation projects 2008DFA20420 of the
Ministry of Science and Technology of China.  The PLATO observatory
was supported by the Australian Research Council and the Australian
Antarctic Division.  Iridium communications were provided by the US
National Science Foundation and the United States Antarctic
Program. The authors wish to thank all members of the 2008 and 2009
PRIC Dome A expeditions for their heroic effort in reaching the site
and for providing invaluable assistance to the expedition astronomers
in setting up and servicing the PLATO observatory and its associated
instrument suite. Additional financial contributions have been made by
the institutions involved in this collaboration.

\end{document}